%% file: EXO-19-007_temp.tex
\pdfoutput=1

\documentclass[11pt,twoside,a4paper,cmspaper,final,collab]{cms-tdr}
\def\svnVersion{db925f7-D}\def\svnDate{2019/10/03}\def\cmsCernNoTag{CERN-EP-2019-159}\def\cmsCernDate{\today}\def\cmsMessage{"Published in the Journal of High Energy Physics as \href{http://dx.doi.org/10.1007/JHEP10(2019)139}{\doi{10.1007/JHEP10(2019)139}.}"}

\begin{document}\cmsNoteHeader{EXO-19-007}

\hyphenation{had-ron-i-za-tion}
\hyphenation{cal-or-i-me-ter}
\hyphenation{de-vices}
\newcommand{\vg}{\ensuremath{\vec{p}^{\gamma}_\mathrm{T}}\xspace}
\newcommand{\ptg}{\ensuremath{p^{\gamma}_\mathrm{T}}\xspace}
\newcommand{\vll}{\ensuremath{\overrightarrow{\ell\ell}}\xspace}
\newcommand{\vmetg}{\ensuremath{p_\mathrm{T}}^{\ptvecmiss+\vg}\xspace}
\newcommand{\mll}{\ensuremath{m_{\ell\ell}}\xspace}
\newcommand{\absetag}{\ensuremath{\abs{\eta^\gamma}\xspace}}

\newlength\cmsTabSkip\setlength{\cmsTabSkip}{1ex}

\cmsNoteHeader{EXO-19-007}

\title{Search for dark photons in decays of Higgs bosons produced in association with $\cPZ$ bosons in proton-proton collisions at $\sqrt{s}=13\TeV$}

\author*[inst1]{CMS experiment}

\date{\today}

\abstract{
A search is presented for a Higgs boson that is produced in association with a $\cPZ$ 
boson and that decays to an undetected particle together with an isolated photon. 
The search is performed by the CMS Collaboration at the Large Hadron Collider 
using a data set corresponding to an integrated luminosity of $137\fbinv$ 
recorded at a center-of-mass energy of $13\TeV$. 
No significant excess of events above the expectation from the standard model 
background is found. The results are interpreted in the context of a theoretical 
model in which the undetected particle is a massless dark photon. 
An upper limit is set on the product of the cross section for associated Higgs 
and $\cPZ$ boson production and the branching fraction for such a Higgs boson decay, 
as a function of the Higgs boson mass. For a mass of  125\GeV, 
assuming the standard model production cross section, this corresponds to 
an observed (expected) upper limit on this branching fraction of 4.6 (3.6)\% at 95\% confidence level. 
These are the first limits on Higgs boson decays to final states that include an undetected massless dark photon.}

\hypersetup{%
pdfauthor={CMS Collaboration},%
pdftitle={Search for dark photons in decays of Higgs bosons produced in association with Z bosons in proton-proton collisions at sqrts = 13 TeV},%
pdfsubject={CMS},%
pdfkeywords={CMS, physics, Higgs boson, exotic decays}}

\maketitle
\section{Introduction}
\label{sec:Introduction}

Following the discovery of a Higgs
boson by the ATLAS and CMS Collaborations~\cite{AtlasPaperCombination, CMSPaperCombination, CMSPaperCombination2}, 
a primary focus of the LHC physics program has been the study of the properties of the new particle.
The observation of a sizable branching fraction of the Higgs boson to invisible 
or almost invisible final states~\cite{Ghosh:2012ep,Martin:1999qf,Bai:2011wz,Gori} would be 
a strong sign of physics beyond the standard model (BSM). 
Studies of the new boson at a mass of about 125\GeV~\cite{Aad:2015zhl,Sirunyan:2017exp} show no significant 
deviation from the standard model (SM) Higgs boson hypothesis, and measurements of its couplings 
constrain its partial decay width to undetected decay modes~\cite{Khachatryan:2016vau,Sirunyan:2018koj}. 
Assuming that the couplings of the Higgs boson to $\PW$ and $\cPZ$ bosons are smaller than
the SM values, an upper limit of 38\% has been obtained at 95\% confidence level ($\CL$) on the branching fraction of the  125\GeV Higgs boson to BSM 
particles~\cite{Sirunyan:2018koj}.

\begin{figure}[htbp]
\centering
\includegraphics[width=0.48\textwidth]{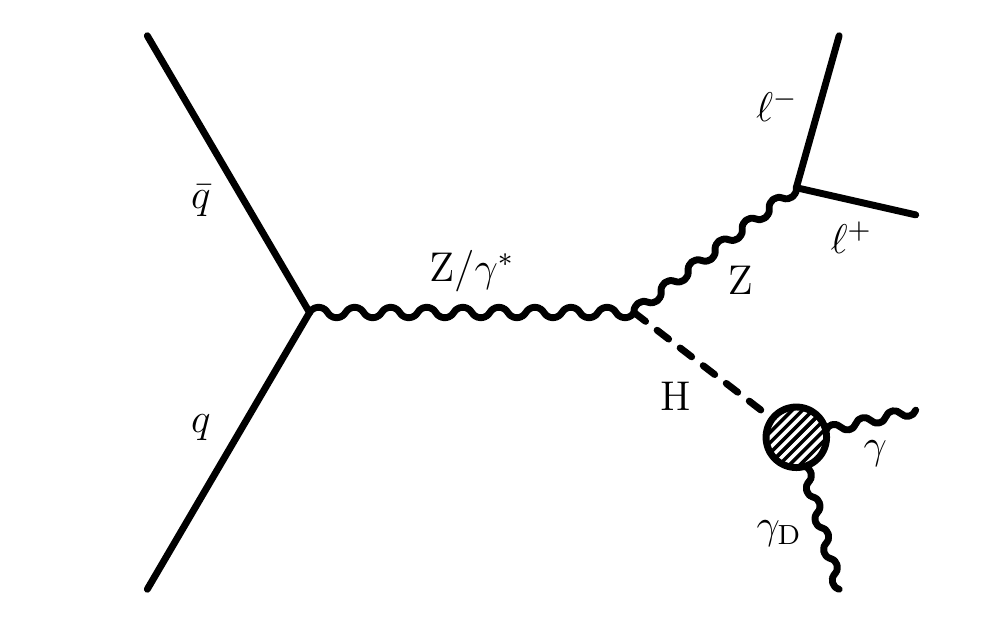}
\caption{A Feynman diagram for the production of the 
$\cPZ(\to \ell\ell) \PH(\to \gamma \gamma_\mathrm{D})$ final state.\label{fig:exo-higgs-decay}}
\end{figure}

This paper presents a search for a scalar boson $\PH$ produced in association 
with a $\cPZ$ boson and decaying to an undetected particle together with a photon. 
Several BSM models predict Higgs boson decays to undetected 
particles and photons~\cite{Gori,Djouadi:1997gw,Petersson:2012dp}. In this search, 
the target final state is $\cPZ(\to \ell\ell) \PH(\to \gamma \gamma_\mathrm{D})$, 
where $\ell = \Pe, \mu$, and $\gamma_\mathrm{D}$ is a massless dark photon that couples to the Higgs boson through a 
charged dark sector~\cite{Gabrielli:2013jka,Gabrielli:2014oya,Biswas:2016jsh,Biswas:2017anm}, 
and is undetected in the CMS detector. 
The branching fraction to such an invisible particle and a photon, 
$\mathcal{B}(\PH \to \text{invisible}+\gamma)$, can be as large as 5\%, and be consistent 
with all model parameters and current LHC constraints~\cite{Gabrielli:2014oya}. 
A Feynman diagram for such a process is shown in Fig.~\ref{fig:exo-higgs-decay}. 
While the main focus is the case where the production cross section ($\sigma_{\cPZ\PH}$) 
is assumed to be the same as that for the SM-like Higgs boson with a mass of 125\GeV, the same
analysis is also used to search for heavy neutral Higgs bosons with masses between 125 and 300\GeV, since similar decays are also possible for potential non-SM scalar bosons.

In the SM, a similar signature to the signal process arises when the Higgs boson decays via 
$\PH\to \cPZ\gamma \to \PGn\PAGn\gamma$, which has a
branching fraction of $3 \times 10^{-4}$. 
Searches for the decay $\PH \to \cPZ\gamma$ using $\cPZ \to \ell\ell$ final states have yielded 
an upper limit at 95\% $\CL$ on the product of the cross section and branching fraction of 
about four times the SM expectation~\cite{Sirunyan:2018tbk,Aaboud:2017uhw}. With the available data set, 
the present search  is not sensitive to this SM decay, but because of enhancements that may arise from BSM 
physics, the search may be sensitive to Higgs boson decays to 
invisible particles and photons. 
The analysis summarized in this paper uses proton-proton ($\Pp\Pp$) collision 
data collected at $\sqrt{s} = 13\TeV$ by the CMS detector in 2016--18 with a 
total integrated luminosity of $137\fbinv$. 
A similar search was performed by the CMS Collaboration using the data collected 
at $\sqrt{s} = 8\TeV$~\cite{Khachatryan:2015vta}, although that analysis investigated 
Higgs bosons produced both in gluon-gluon fusion and in association with a $\cPZ$ boson.

The main backgrounds in this analysis arise from $\PW\cPZ$ and $\cPZ\cPZ$ 
production, where an electron is mis-identified as a photon, or where additional leptons are not identified 
because they fail either the lepton identification criteria or the kinematic selections. A second 
set of backgrounds are due to $\PW\PW$ and top quark production, where the invariant mass of the 
lepton pair falls into the $\cPZ$ boson mass window. 
There are also small contributions from other multiboson production processes, such as $\cPZ\gamma$. 
To enhance the discrimination between the potential signal and the remaining background processes, 
a binned maximum-likelihood fit to several signal and control regions is performed. 

\section{The CMS detector}
\label{sec:cms}

The central feature of the CMS apparatus is a superconducting solenoid of 
6\unit{m} internal diameter, providing a magnetic field of 3.8\unit{T}. Within 
the solenoid volume are a silicon pixel and strip tracker, a lead tungstate 
crystal electromagnetic calorimeter (ECAL), and a brass and scintillator hadron 
calorimeter, each composed of a barrel and two endcap sections. Forward 
calorimeters extend the pseudorapidity ($\eta$) coverage provided by the barrel 
and endcap detectors. Muons are detected in gas-ionization chambers embedded in 
the steel magnetic flux-return yoke outside the solenoid. A more detailed description 
of the CMS detector, together with a definition of the coordinate system used and 
the relevant kinematic variables, can be found in Ref.~\cite{Chatrchyan:2008aa}. 
Events of interest are selected using a two-tiered trigger 
system~\cite{Khachatryan:2016bia}. The first level, composed of custom hardware 
processors, uses information from the calorimeters and muon detectors to select 
events, while the second level selects events by running a version of the full 
event reconstruction software optimized for fast processing on a farm of computer processors.

\section{Data samples and event reconstruction}
\label{sec:objects}

The data used in this search were collected in three separate LHC operating periods 
in 2016, 2017, and 2018. 
The three data sets are analyzed independently, with appropriate calibrations and corrections 
to take into account the different LHC running conditions and CMS detector performance.

Monte Carlo (MC) simulated events are used to model the expected signal and background 
yields. Three sets of simulated events for each process are 
needed to match the different data taking conditions in the three 
different years. 
The next-to-leading-order (NLO) $\POWHEG$ v2~\cite{Frixione:2002ik,Nason:2004rx,Frixione:2007vw,Alioli:2008gx,Alioli:2010xd} 
generator is used to simulate the $\cPZ\PH$ signal process at NLO in quantum chromodynamics (QCD), as well as the $\ttbar$, $\PQt\PW$, 
and diboson processes. The BSM Higgs boson production cross sections as a function of the 
Higgs boson mass ($m_{\PH}$) for the $\cPZ\PH$ process are taken from Refs.~\cite{Heinemeyer:2013tqa,deFlorian:2016spz}. 
The signal samples are generated for masses of 125, 200, and 300\GeV. 
Production of $\ttbar\PW$, $\ttbar\cPZ$, $\ttbar\gamma$, and 
triple vector boson (\PV\PV\PV) events is generated at NLO 
in QCD using the $\MGvATNLO$ 2.2.2 (2.4.2) generator for 2016 (2017 and 2018) \cite{Alwall:2014hca,MadGraph,Frederix2012} samples. 
The NNPDF 3.0 NLO \cite{Ball:2014uwa} 
(NNPDF 3.1 next-to-next-to-leading-order~\cite{Ball:2017nwa}) parton distribution functions 
(PDFs) are used for simulating all 2016 (2017 and 2018) samples. 
For all processes, the parton showering and hadronization are simulated using 
$\PYTHIA$ 8.226 (8.230) in 2016 (2017 and 2018)~\cite{Sjostrand:2007gs}.
The modeling of the underlying event is generated using the 
CUETP8M1~\cite{Skands:2014pea,Khachatryan:2015pea} and CP5 tunes~\cite{Sirunyan:2019dfx} 
for simulated samples corresponding to the 2016 and 2017--18 data sets, respectively.

All MC simulation events are processed through a simulation of the CMS detector based on 
\GEANTfour~\cite{Geant} and are reconstructed with the same algorithms as 
used for data. Additional $\Pp\Pp$ interactions in the same and nearby bunch 
crossings, referred to as pileup, are also simulated. 
The distribution of the number of such interactions in the simulation is adjusted 
to match the one observed in the data. 
The average number of pileup interactions was 23 (32) in 2016 (2017 and 2018).

For this search, collision events were collected using single-electron and single-muon
triggers that require the presence of an isolated lepton with transverse momentum ($\pt$) larger than 
24 and 27\GeV, respectively. In addition, a set of dilepton triggers with lower $\pt$ thresholds 
were used, ensuring a trigger efficiency above 99\% for events that would satisfy the subsequent offline selection.

Information from all subdetectors is combined and used by the CMS particle-flow (PF) 
algorithm~\cite{Sirunyan:2017ulk} for particle reconstruction and identification.
Jets are reconstructed by clustering PF candidates using the anti-\kt 
algorithm~\cite{Cacciari:2008gp} with a distance parameter 0.4. 
Jets are calibrated in the simulation, and separately in data, accounting for energy 
deposits of neutral particles from pileup and any nonlinear detector response~\cite{Khachatryan:2016kdb}. 
Jets with $\pt > 30\GeV$ and $\abs{\eta} < 4.7$ are considered in the analysis. 
The effect of pileup is mitigated through a charged-hadron subtraction technique, 
which removes the energy of charged hadrons not originating from the primary 
interaction vertex (PV)~\cite{CMS-PAS-JME-14-001}. 
The PV is defined as the vertex with the largest value of summed 
physics-object $\pt^2$. Here, the physics objects are the jets clustered using the jet finding 
algorithm~\cite{Cacciari:2008gp,Cacciari:2011ma} with the tracks assigned to the 
vertex as inputs, and the associated missing transverse momentum, taken as the 
negative vector $\pt$ sum of those jets. 

Events are discarded if they contain a jet with $\pt > 20\GeV$ and $\abs{\eta}<2.4$ that is consistent 
with the fragmentation of a \PQb\ quark. The combined secondary vertex 
(CSVv2) \PQb\ tagging algorithm~\cite{Sirunyan:2017ezt} is used for the 2016 data set, while the 
DeepCSV algorithm~\cite{Sirunyan:2017ezt} is used for the 2017 and 2018 data sets. 
For the chosen working points, the efficiency to select \PQb\ quark jets is about 62 (72)\% for CSVv2 (DeepCSV) and 
the rate for incorrectly \PQb\ tagging jets originating from the hadronization of gluons or $\cPqu$, $\cPqd$, $\cPqs$ quarks is about 1\%.

The vector $\ptvecmiss$ is defined as the negative vector $\pt$ sum of all PF 
particle candidates. The magnitude of $\ptvecmiss$ is the missing transverse momentum $\ptmiss$. 
Corrections to jet energies due to detector 
response are propagated to $\ptmiss$~\cite{Khachatryan:2014gga}. Events with 
possible contributions from beam halo processes or anomalous signals in the calorimeters are 
rejected using dedicated filters~\cite{Khachatryan:2014gga}. 

Electrons and muons are reconstructed by associating a track reconstructed 
in the tracking detectors with either a cluster of energy in the 
ECAL~\cite{Khachatryan:2015hwa} or a track in the muon system~\cite{Sirunyan:2018fpa}. 
Electron and muon candidates must pass certain identification criteria to be further 
selected in the analysis. For the ``loose" identification, they must satisfy $\pt > 10\GeV$ and 
$\abs{\eta} < 2.5$ (2.4) for electrons (muons). At the final stage of the lepton selection, 
the medium working points, following the definitions provided in Ref.~\cite{Khachatryan:2015hwa,Sirunyan:2018fpa}, 
are chosen for the identification criteria, including requirements on the impact parameter 
of the candidates with respect to the PV and their isolation with 
respect to other particles in the event~\cite{Sirunyan:2018egh}.

Finally, photon candidates are reconstructed from energy deposits in the 
ECAL~\cite{Khachatryan:2015iwa} within $\abs{\eta} < 2.5$. The identification of the candidates is based 
on shower shape and isolation variables, and the medium working point, 
as described in Ref.~\cite{Khachatryan:2015iwa}, is chosen to select those candidates. In addition, a standard 
``conversion-safe electron veto"~\cite{Khachatryan:2015iwa} is applied to 
reject electrons misidentified as photons.

\section{Event selection}
\label{sec:selection}

The signal topology consists of two oppositely 
charged same-flavor high $\pt$ isolated leptons, electrons or muons, compatible with a $\cPZ$ 
boson decay, large $\ptmiss$, an isolated high $\pt$ photon, and little jet activity. 
The signal cross section is several orders of magnitude lower than that of the major reducible 
background processes, and therefore a stringent selection is required to obtain a sample of 
sufficient purity. To be consistent with the expected topology, 
the selection requires a leading (subleading) lepton with 
$\pt > 25$ (20)\GeV and at least one photon with transverse momentum $\ptg > 25\GeV$. 
To reduce background processes where the lepton pair is not from the decay of a $\cPZ$ boson, 
the dilepton mass must be compatible with that of a $\cPZ$ boson within 15\GeV of the 
pole mass $m_{\cPZ}$~\cite{Tanabashi:2018oca}. 
For the purpose of rejecting the bulk of the $\cPZ\gamma$  background as well as processes 
with little or moderate boost, a $\ptmiss$ greater than 110\GeV and a transverse momentum 
of the dilepton system $\pt^{\ell\ell}$ larger than 60\GeV are required.

To reduce the background from $\PW\cPZ$ events with a third lepton from 
the $\PW$ boson decay, events are removed if, in addition to the two leptons satisfying the full selection criteria, 
there are any loosely identified leptons. To suppress the top quark background, 
events are rejected if any jet passes the \cPqb\ tagging selection (\cPqb\ jet veto) 
described above, or if there are more than two identified jets in the event.

The signal topology is characterized by a dilepton system ($\vll$) with large 
$\pt$ balanced in the transverse plane by the $\ptvecmiss+\vg$ system from the Higgs boson decay. Therefore, to reject most of the 
background from $\cPZ\gamma$ events with misreconstructed $\ptmiss$, 
the azimuthal angle between the $\vll$ and $\ptvecmiss+\vg$ systems ($\Delta \phi_{\vll,\ptvecmiss+\vg}$) is required to be greater 
than 2.5 rad, 
the quantity $\abs{\vmetg-\pt^{\ell\ell}}/\pt^{\ell\ell}$ is required 
to be smaller than 0.4, and the azimuthal angle between the leading jet and $\ptvecmiss$ 
($\Delta \phi_{\mathrm{jet},\ptvecmiss}$) should be greater than 0.5 rad. In addition, 
the mass of the dilepton and photon system ($m_{\ell\ell\gamma}$) must 
be greater than  100\GeV to reject resonant $\cPZ\gamma$ events, where the photon originates from final-state radiation. 
Finally, the transverse mass of the $\ptvecmiss$ and photon 
system, defined as $\mT \equiv \sqrt{\smash[b]{2\ptmiss\ptg[1-\cos(\Delta \phi_{\ptvecmiss,\vg})]}}$, 
must be smaller than  350\GeV, which rejects events where the dilepton 
and photon objects are weakly correlated, or where the photon momentum is mismeasured. The quantity 
$\Delta \phi_{\ptvecmiss,\vg}$ is the azimuthal angle between $\ptvecmiss$ and the photon. 
A summary of the selection  for the analysis is shown in Table~\ref{tab:selectioncuts}. 

\begin{table}[htbp]
  \centering
  \topcaption{Summary of the selection criteria and the main background processes.}
  \label{tab:selectioncuts}
  \begin{tabular} {lcc}
\hline
  Variable & Selection & Reject\\
  \hline
Number of leptons                                              & Exactly 2 leptons, $\pt>25/20\GeV$	 & $\PW\cPZ$, $\cPZ\cPZ$, \PV\PV\PV \\
Number of photons                                              & $\geq$1 photon, $\ptg>25\GeV$ 	         & All but $\cPZ\gamma$             \\
$\abs{\mll - m_{\cPZ}}$                                        & $<$15\GeV                               & $\PW\PW$, Top quark              \\
$\ptmiss$                                                      & $>$110\GeV                              & $\cPZ\gamma$  	            \\
$\pt^{\ell\ell}$                                               & $>$60\GeV                               & $\cPZ\gamma$  	            \\
$\cPqb$ jet veto                                               & Applied                                 & Top quark, \PV\PV\PV	            \\
Jet counting                                                   & $\leq$2                                 & Top quark, \PV\PV\PV	            \\
$\Delta \phi_{\vll,\ptvecmiss+\vg}$                            & $>$2.5 rad 			         & $\cPZ\gamma$  	            \\
$\abs{\vmetg-\pt^{\ell\ell}}/\pt^{\ell\ell}$                   & $<$0.4 			         & $\cPZ\gamma$  	            \\
$\Delta \phi_{\text{jet},\ptvecmiss}$                        & $>$0.5 rad 			         & $\cPZ\gamma$  	            \\
$m_{\ell\ell\gamma}$                                           & $>$100\GeV                              & $\cPZ\gamma$  	            \\
$\mT$                                                          & $<$350\GeV                              & $\PW\PW$, Top quark              \\
  \hline
  \end{tabular}
\end{table}

\section{Background estimation}
\label{sec:backgrounds}

A combination of methods based on control samples in data and simulation is used to 
estimate background contributions. 
Uncertainties related to the theoretical and experimental predictions are 
taken into account, as described in Section~\ref{sec:systematics}. 
Background contributions are categorized depending on whether they produce 
at least one lepton pair from the decay of a $\cPZ$ boson (resonant contributions) 
or no such lepton pair (nonresonant contributions). 
The expected yield for the irreducible background from 
$\Pp\Pp \to \cPZ(\to \ell\ell)\PH(\to\cPZ\gamma)\to\ell\ell\PGn\PAGn\gamma$
is less than 0.1 events and is consequently ignored in the analysis.

\subsection{Nonresonant dilepton backgrounds}

The contribution from the nonresonant dilepton backgrounds, mostly $\PW\PW$ and top quark processes, 
is estimated by exploiting the lepton flavor symmetry in the final states of
these processes~\cite{Sirunyan:2017qfc}. 
A control region based on the $\Pe^{\pm}\mu^{\mp}$  final state, whose branching 
fraction is twice that of either same flavor lepton pair final state, is 
used to estimate these backgrounds in the $\Pep\Pem$ and $\PGmp\PGmm$ channels.
This region is completely dominated by this nonresonant dilepton background. 
The method considers the differences between the electron and muon
identification efficiencies when extrapolating from the different-flavor to the same-flavor final states. 
The resulting predictions agree with the number of background events estimated by applying 
the same method to the simulation. The chosen $\Pe\mu$ control region contains 3 events 
that satisfy the full analysis selection, to be compared with an expectation of 
$2.8 \pm 0.5 \stat$ from the simulation.

\subsection{Resonant background with genuine missing transverse momentum}

The resonant background with genuine missing transverse momentum in which an electron is mis-identified as a photon is 
dominated by the $\PW\cPZ \to \Pe\PGn\ell\ell$ process. 
In this case, the background comes from events where the electron 
from the $\PW$ boson decay is wrongly identified as a photon. 
The electron to photon misidentification rate is measured in $\cPZ \to \Pe\Pe$ events 
by comparing the ratio of $\Pe\gamma$ to $\Pe\Pe$ pairs consistent with 
the $\cPZ$ boson mass, as reconstructed in data and simulation. 
The average misidentification rate is 1--5\%, with the larger values corresponding to higher 
photon pseudorapidity \absetag. 

Background processes with two leptons and a genuine hard photon are estimated 
with the simulation. These events arise primarily from the 
$\PW\cPZ\to\ell\PGn\ell\ell$ process (where the lepton from the 
$\PW$ boson decay is not identified) and $\cPZ\cPZ \to 2\ell 2\PGn$. 
In both cases an additional hard photon must be radiated.

To assess the normalization of the $\PW\cPZ \to \ell\PGn\ell\ell$ background, a 
control region is selected in data by applying the full selection on events 
where the selected lepton from the $\PW$ boson decay plays the role of the photon. 
In this region, 231 events are observed in data, while the simulation 
predicts $241 \pm 4 \stat$ events.

\subsection{Resonant background with no genuine missing transverse momentum}

The background from $\cPZ\gamma$ or $\cPZ+\text{jets}$ events is predicted by 
the simulation to be less than 5\% of the total background, because of the stringent 
selection used. 
One of the data control regions used to verify that the background is correctly 
estimated selects events with lower $\ptmiss$ than the default selection. 
Within the uncertainties, good agreement between data 
and simulation is found. 
To estimate the overall $\cPZ\cPZ$ normalization, and also to emulate the 
$\cPZ\gamma$ process, a four-lepton sample is selected in data, and the full analysis selection 
is performed, with one of the $\cPZ$ boson dilepton pairs as a photon. 
In this control region, $5.1 \pm 0.2 \stat$ events are expected from simulation, 
while 7 events are observed in data.

\section{Signal extraction}
\label{sec:fitting}

After applying the event selection, the 2016, 2017, and 2018 data sets are treated individually 
in order to maximize the sensitivity of the combination, since the signal-to-background ratio is 
different in each case. On the other hand, the electron and muon channels are 
merged because they show a similar signal-to-background ratio.

To discriminate between the potential signal and the remaining background processes, 
a binned maximum-likelihood fit to the $\mT$ spectrum is performed. 
The signal spectrum shows a Jacobian peak with an end-point at $\mT \sim m_{\PH}$, while the 
background processes have either a flat distribution or display an increase towards 
lower values of $\mT$. Since the contamination from electrons misidentified as photons 
is larger at large \absetag\ values, improved sensitivity is achieved  by considering 
separately events with the selected photon at low- or high-\absetag.

In the maximum-likelihood fit, each bin of the $\mT$ distribution is separated into 
a low-\absetag\ ($\absetag < 1$) and a high-\absetag\ ($\absetag > 1$) bin, 
for the signal region and the $\Pe\mu$, $\PW\cPZ$, and $\cPZ\cPZ$ control regions. 
For each individual bin, a Poissonian likelihood
term is used to describe the fluctuation of the yields around the expected
central value, which is given by the sum of the contributions from signal 
and background processes. 
The uncertainties affect the overall normalizations of the signal and background yields, 
as well as the shapes of the predictions across the distributions of the observables.
Correlations between systematic uncertainties in different categories
are taken into account. Uncertainties that purely affect the normalization within a category are 
incorporated as nuisance parameters with log-normal priors, while those associated with changes 
in shapes are assigned probability density functions described by a polynomial interpolation 
with a Gaussian constraint. The total likelihood is defined as the product of the 
likelihoods of the individual bins and the probability density functions 
for the nuisance parameters, including the product of the likelihood for the individual years. 
In summary, the maximum-likelihood fit function, $\mathcal{L}$, can be written as:
\begin{linenomath}
\begin{equation}
\begin{split}
 \mathcal{L} = \prod_{i,j,k} & \mathcal{P} \Big( N^{\mathrm{SR}}_{\mathrm{Obs},(i,j,k)} \mid N^{SR}_{\mathrm{Other,(i,j,k)}}(\theta) + \mu N^{SR}_{\cPZ\PH,(i,j,k)}(\theta) + \mu^{\mathrm{Nonres}} N^{SR}_{\mathrm{Nonres,(i,j,k)}}(\theta) \\
 & + \mu^{\PW\cPZ} N^{SR}_{\PW\cPZ,(i,j,k)}(\theta) + \mu^{\cPZ\cPZ} N^{SR}_{\cPZ\cPZ,(i,j,k)}(\theta) \Big)										 \\ 
 & \mathcal{P} \Big( N^{\Pe\mu}_{\mathrm{Obs},(i,k)} \mid N^{\Pe\mu}_{\mathrm{Other,(i,k)}}(\theta) + \mu^{\mathrm{Nonres}}N^{\Pe\mu}_{\mathrm{Nonres,(i,k)}}(\theta) \Big) \\ 
 & \mathcal{P} \Big( N^{3\ell}_{\mathrm{Obs},(i,k)} \mid N^{3\ell}_{\mathrm{Other,(i,k)}}(\theta) + \mu^{\PW\cPZ} N^{3\ell}_{\PW\cPZ,(i,k)}(\theta) \Big)			  \\ 
 & \mathcal{P} \Big( N^{4\ell}_{\mathrm{Obs},(i,k)} \mid N^{4\ell}_{\mathrm{Other,(i,k)}}(\theta) + \mu^{\cPZ\cPZ} N^{4\ell}_{\cPZ\cPZ,(i,k)}(\theta) \Big)				  \\ 
 & \re^{-(\theta-\hat{\theta})^{2}/2},																	  \\
\end{split}
\end{equation}
\end{linenomath}
where $i$ runs over the three data-taking periods, $j$ runs over the $\mT$ bins, $k$ runs over the two \absetag\ values, 
$\mathcal{P}(N\mid \lambda)$ is the Poisson probability, 
$\theta$ are nuisance parameters for the systematic uncertainties, and 
$\hat{\theta}$ are their default values. The values $N^{\mathrm{SR}}_{\mathrm{Obs},(i,j,k)}$, $N^{\Pe\mu}_{\mathrm{Obs},(i,k)}$, $N^{3\ell}_{\mathrm{Obs},(i,k)}$, and $N^{4\ell}_{\mathrm{Obs},(i,k)}$ are the observed data events 
in the signal region, and the $\Pe\mu$, $\PW\cPZ$, and $\cPZ\cPZ$ control regions, respectively. 
The parameters $\mu$, $\mu^{\mathrm{Nonres}}$, $\mu^{\PW\cPZ}$, and $\mu^{\cPZ\cPZ}$ are the fit normalization 
factors for the signal, nonresonant, $\PW\cPZ$, and $\cPZ\cPZ$ processes, respectively. The values 
$N_{\cPZ\PH}$, $N_{\mathrm{Nonres}}$, $N_{\PW\cPZ}$, $N_{\cPZ\cPZ}$, and $N_{\mathrm{Other}}$ are the 
expected number of events for the signal, nonresonant, $\PW$, $\cPZ\cPZ$, and remaining processes, respectively, 
for the different regions. This approach follows that of Ref.~\cite{LHC-HCG-Report}, 
where more details can be found.

\begin{figure}[htbp]
\centering
\includegraphics[width=0.48\textwidth]{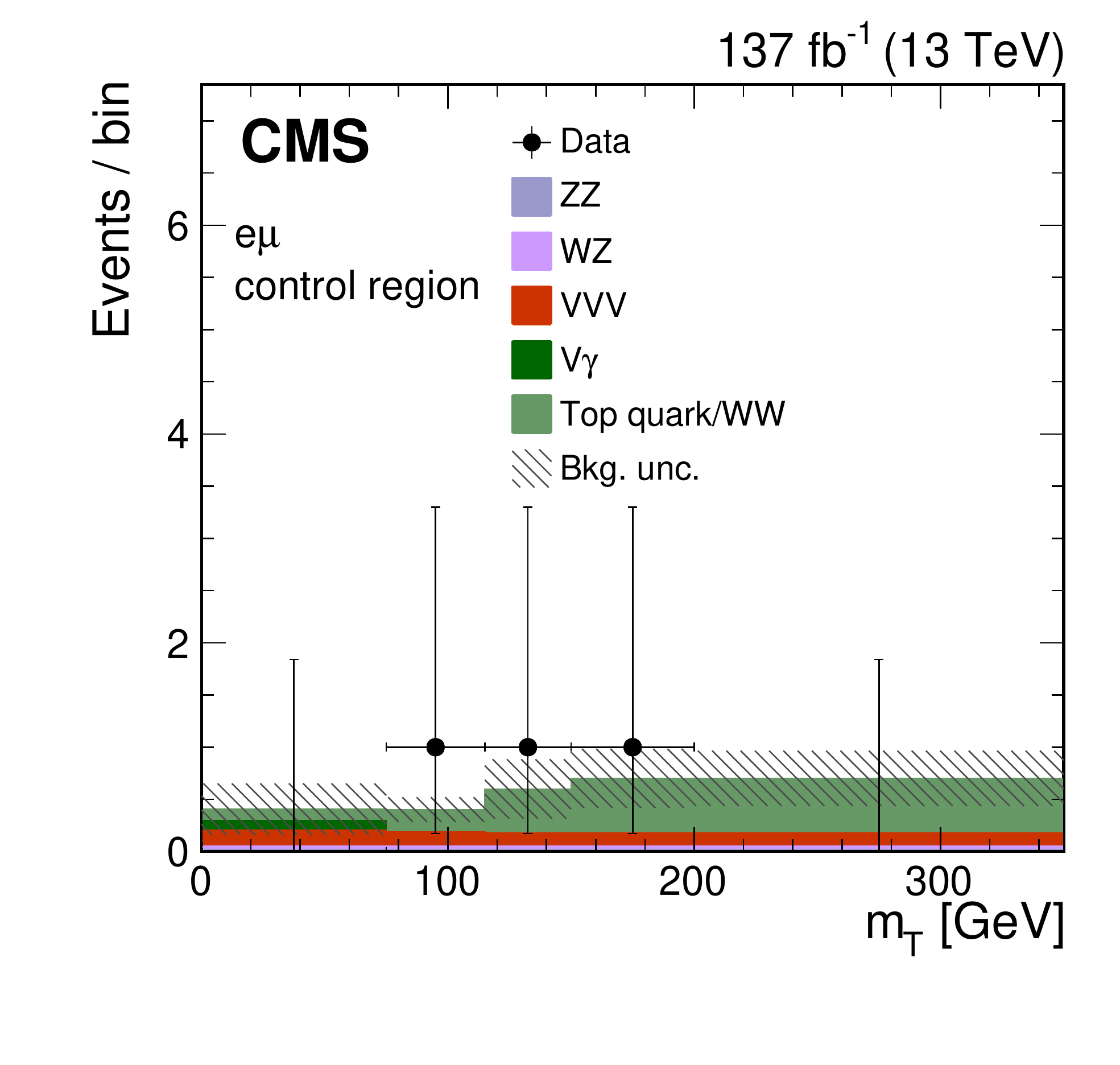}
\includegraphics[width=0.48\textwidth]{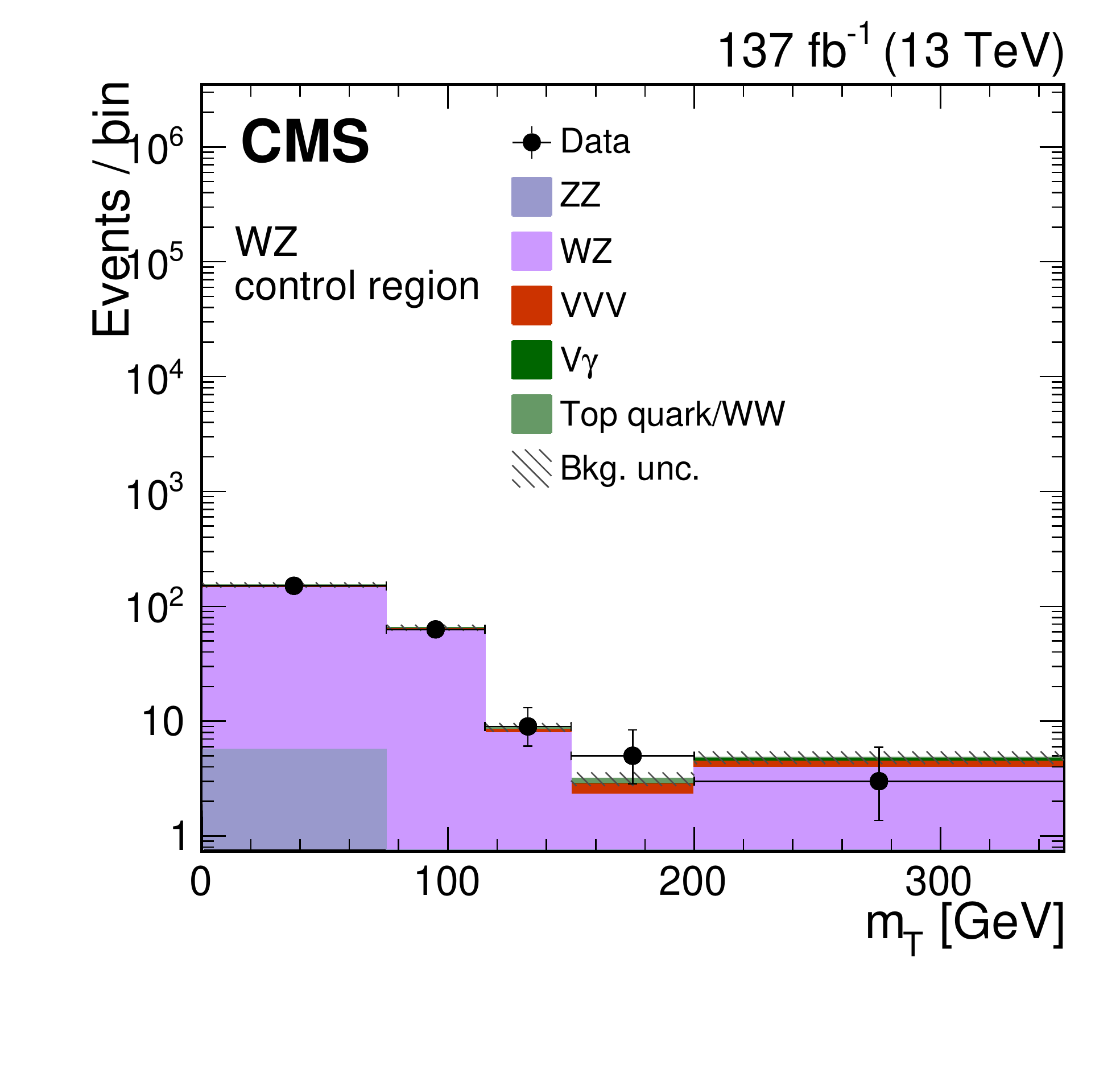}
\includegraphics[width=0.48\textwidth]{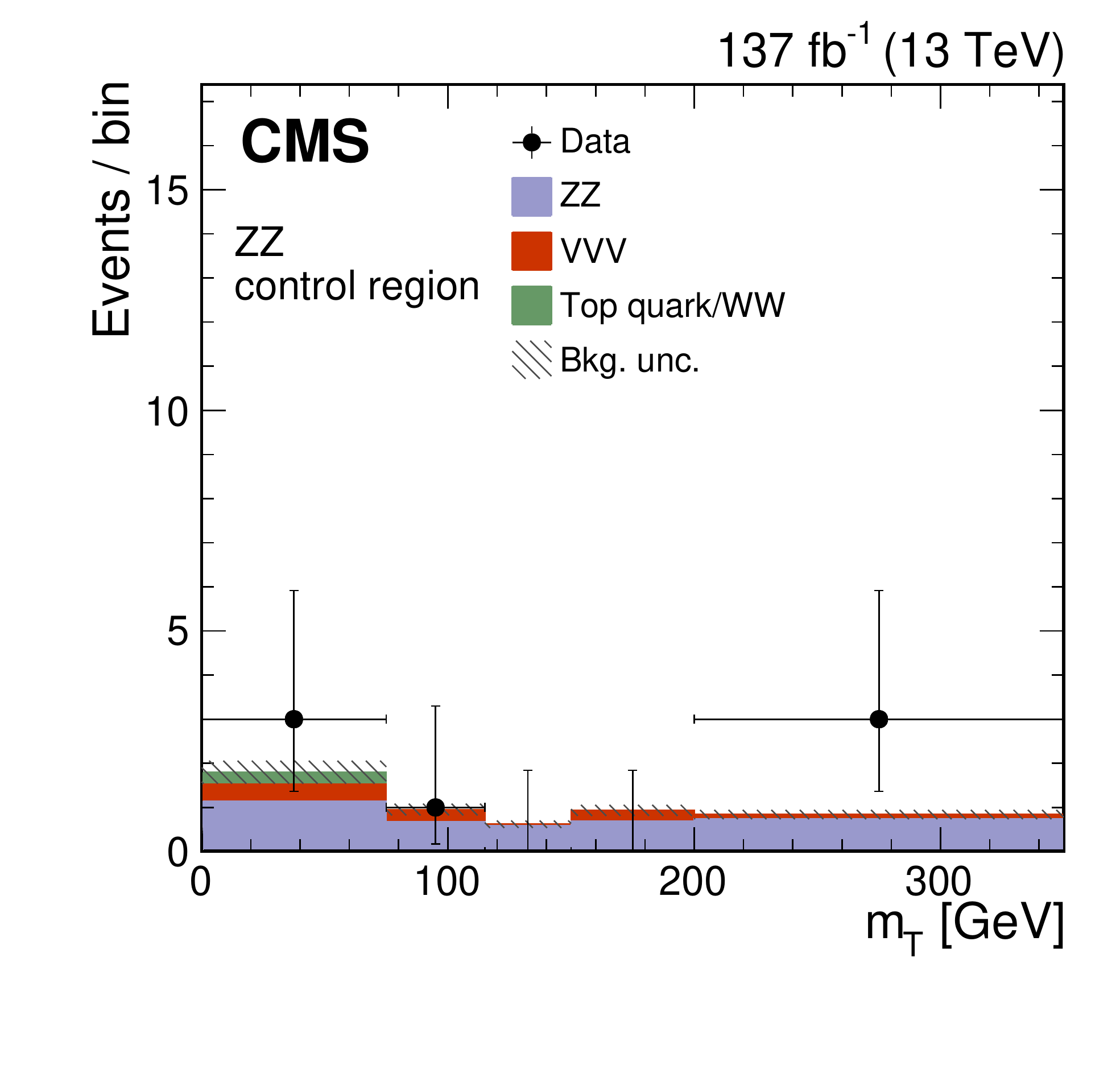}
\caption{The $\mT$ distributions for the $\Pe\mu$, $\PW\cPZ$, and $\cPZ\cPZ$ control regions after 
the simultaneous fit to data in the signal and control regions. Statistical and systematic 
uncertainties in the expected background yields are represented by the hatched band. 
Vertical bars represent data statistical uncertainties, while 
horizontal bars represent the bin widths.\label{fig:zhg_control_regions}}
\end{figure}

The $\mT$ distributions for the $\Pe\mu$, $\PW\cPZ$, and $\cPZ\cPZ$ control regions 
are shown in Fig.~\ref{fig:zhg_control_regions}. 
This analysis fits the $\mT$ distributions for two regions of \absetag, a procedure 
that improves the expected limits by about 30 to 50\% compared with the 
results from simply counting the contents of a single $\mT$ bin for each 
\absetag\ region, as was done in Ref.~\cite{Khachatryan:2015vta}. The improvement from 
splitting the data into two regions of \absetag\ is about 4\%.

\section{Efficiencies and systematic uncertainties}
\label{sec:systematics}

Several sources of systematic uncertainty are taken into
account in the maximum-likelihood fit. 
For each source of uncertainty, the effects on the final distributions are considered correlated.

The assigned uncertainties in the integrated luminosity measurements for
the data used in this analysis are 2.5, 2.3, and 2.5\% for the 
2016, 2017, and 2018 data 
samples~\cite{CMS-PAS-LUM-17-001, CMS-PAS-LUM-17-004, CMS-PAS-LUM-18-002}, respectively. 
They are treated as uncorrelated across the three data sets. 

The simulation of pileup events assumes a total inelastic pp cross section of 69.2~mb, 
with an associated uncertainty of 5\%~\cite{ATLAS:2016pu,Sirunyan:2018nqx}, which has 
an impact on the expected signal and background yields of about 1\%.

Discrepancies in the lepton and photon reconstruction and identification
efficiencies between data and simulation are corrected by applying 
scale factors to all MC simulation samples.
These scale factors are determined using $\cPZ \to \ell\ell$ 
events in the $\cPZ$ boson peak region that were recorded with 
unbiased triggers~\cite{Sirunyan:2018fpa,Khachatryan:2015hwa}. 
The scale factors depend on the $\pt$ and $\eta$ of the lepton, and are within 2\% of 
unity for both electrons and muons. 
The uncertainty in the determination of the trigger efficiency leads to an uncertainty 
smaller than 1\% in the expected signal yield. The lepton momentum scale uncertainty is computed by varying the 
momenta of the leptons in simulation by their uncertainties, and repeating the 
analysis selection. The resulting yield uncertainties are $\approx$1\% for both electrons and muons. 
The above procedure is applied also to determine the scale factors for photons, 
and the yield uncertainty for photon candidates is $\approx$1.5\%. 
These uncertainties are treated as correlated across the three data sets.

The uncertainty in the calibration of the jet energy scale directly affects 
the acceptance of the jet multiplicity requirement and the $\ptmiss$ measurement. 
These effects are estimated by shifting the jet energy in the simulation up and down by one standard deviation. 
The uncertainty in the jet energy scale is 2--5\%, depending on $\pt$ and $\eta$~\cite{Khachatryan:2016kdb}, 
and the impact on the expected signal and background yields is about 3\%. 

In this analysis, \cPqb\ tagging is used to reject events with genuine $\cPqb$ quark jets in the final 
state, since signal events have no \cPqb\ quarks to first order in the decay channel of interest. 
The \cPqb\ tagging efficiency in the simulation is corrected using scale factors determined 
from data~\cite{Sirunyan:2017ezt}. These values are estimated separately for 
correctly and incorrectly identified jets. Each set of values results in the 
\cPqb\ tagging efficiency of about 1--4\%, and the impact on the expected signal and background yields is about 1\%. 
The uncertainties in the jet energy scale and \cPqb\ tagging are treated as 
uncorrelated across the three data sets.

The theoretical uncertainties due to the choice of the 
QCD renormalization and factorization scales are estimated 
by varying these scales independently up and down by a factor of 
two~\cite{Catani:2003zt,Cacciari:2003fi}. 
The variations of the PDF set and the strong coupling constant are used to estimate 
the corresponding uncertainties in the yields of the signal and background processes, 
following Refs.~\cite{Ball:2014uwa,Butterworth:2015oua}. 
The combined impact on the expected yields from the above sources is about 4\%. 
The statistical uncertainty associated with the limited number of simulated events 
is also considered as part of the systematic uncertainty, leading to an impact 
on the expected yields of about 5\%. 
These systematic uncertainties are much smaller than the statistical uncertainty 
because of the limited size of the data sample, and the effect of all systematic uncertainties 
reduces the sensitivity by less than 4\%.

\section{Results}
\label{sec:results}

\begin{figure}[htbp]
\centering
\includegraphics[width=0.48\textwidth]{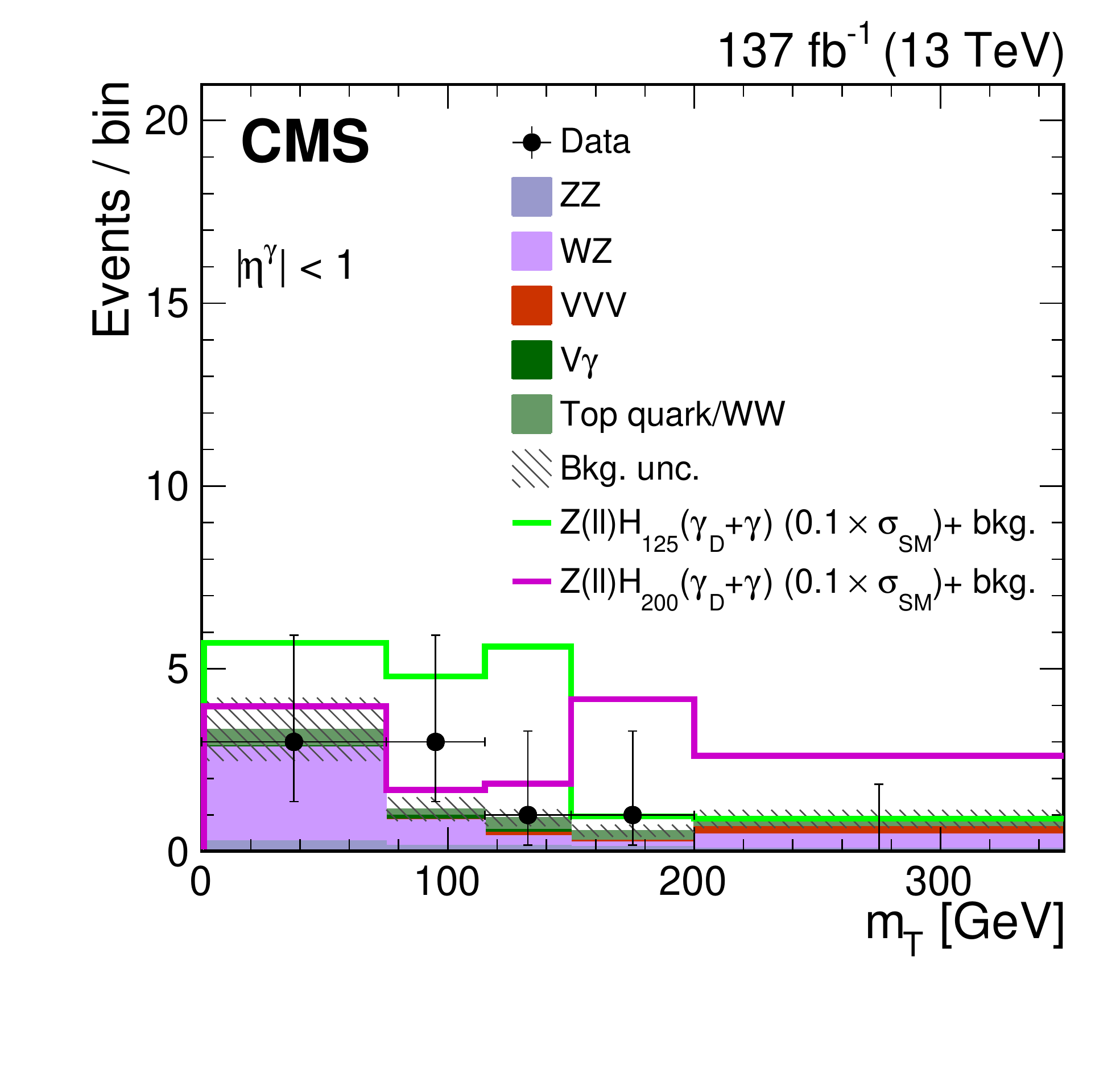}
\includegraphics[width=0.48\textwidth]{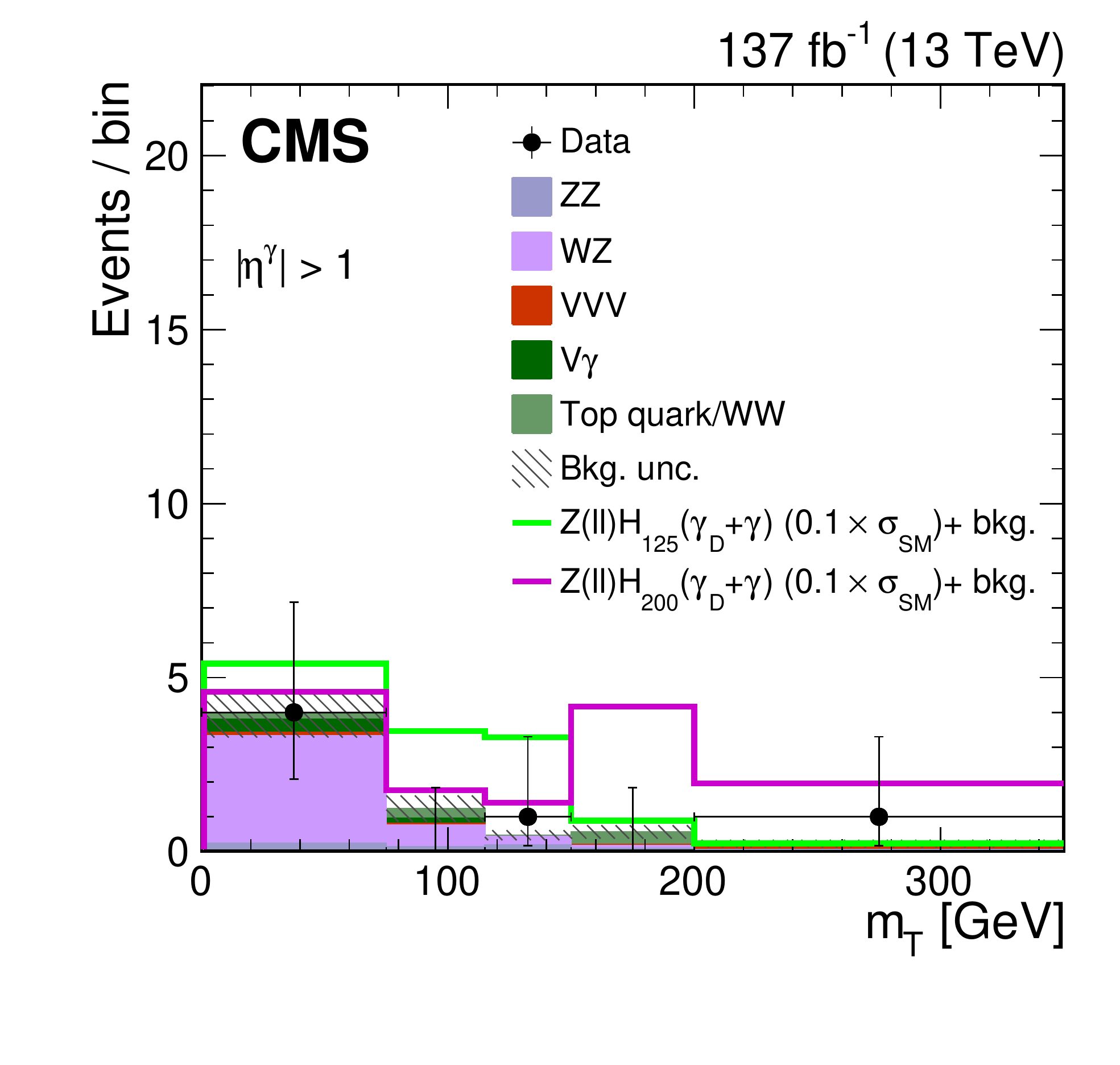}
\caption{The $\mT$ distributions in the signal region for two $m_{\PH}$ values 
for events with $\absetag < 1$ (left) and $\absetag > 1$ (right), after the fit to data. The signal 
size corresponds to $0.1 \sigma_{\cPZ\PH}$ for both values of $m_{\PH}$ shown. 
The signal processes are stacked on top of all backgrounds. 
Statistical and systematic uncertainties in the expected background yields are represented by the hatched band. 
Vertical bars represent data statistical uncertainties, while 
horizontal bars represent the bin widths.\label{fig:finallevel}}
\end{figure}

The numbers of observed and expected events after applying the full selection requirements 
are shown in Table~\ref{tab:zhinvsel}. 
The signal size is chosen for illustration purposes to be $0.1 \sigma_{\cPZ\PH}$, to have 
a rounded number relatively close to the minimum where this analysis is expected to have sensitivity, and to avoid quoting large expected yields. 
The product of acceptance and selection efficiency 
increases at larger $m_{\PH}$ values because of the larger $\pt$ values for all objects in the events. 
The $\mT$ distributions for events with $\absetag < 1$ and $\absetag > 1$ 
after the event selection are shown in Fig.~\ref{fig:finallevel}. Agreement 
between the data and the background-only prediction is observed.

\begin{table}[hbtp]
\centering
\caption{Observed yields, background estimates after the fit to data, 
and signal predictions after the event selection. The signal size corresponds to 
 $0.1 \sigma_{\cPZ\PH}$ for all three $m_{\PH}$ values shown. The combined statistical and systematic uncertainties are reported. 
 The values in parentheses for the signal processes correspond to the products of acceptance and
 selection efficiency for $\cPZ \to \ell\ell$ events.}
\label{tab:zhinvsel}
  \begin{tabular}{lr}
\hline 
Process & Yield \\
\hline
Data                      &    14  \\[\cmsTabSkip]
Nonresonant               &  $ 2.4 \pm 1.1$ \\
$\PW\cPZ$  		  &  $ 8.1 \pm 2.0$ \\
$\cPZ\cPZ$  		  &  $ 1.5 \pm 0.3$ \\
$\cPZ\gamma$  		  &  $ 0.7 \pm 0.7$ \\
Other                     &  $ 0.6 \pm 0.3$ \\[\cmsTabSkip]
Total background          &  $13.3 \pm 3.8$ \\[\cmsTabSkip]
$\cPZ\PH_{125}$ (product of acceptance and efficiency)  &  $17.9 \pm 1.2$ ($ 2.13 \pm 0.14$\%)\\
$\cPZ\PH_{200}$ (product of acceptance and efficiency)  &  $12.3 \pm 0.8$ ($ 6.48 \pm 0.42$\%) \\
$\cPZ\PH_{300}$ (product of acceptance and efficiency)  &  $ 3.9 \pm 0.2$ ($10.20 \pm 0.51$\%) \\
\hline
  \end{tabular}
\end{table}

\begin{figure}[hbtp]
\centering
\includegraphics[width=0.49\textwidth]{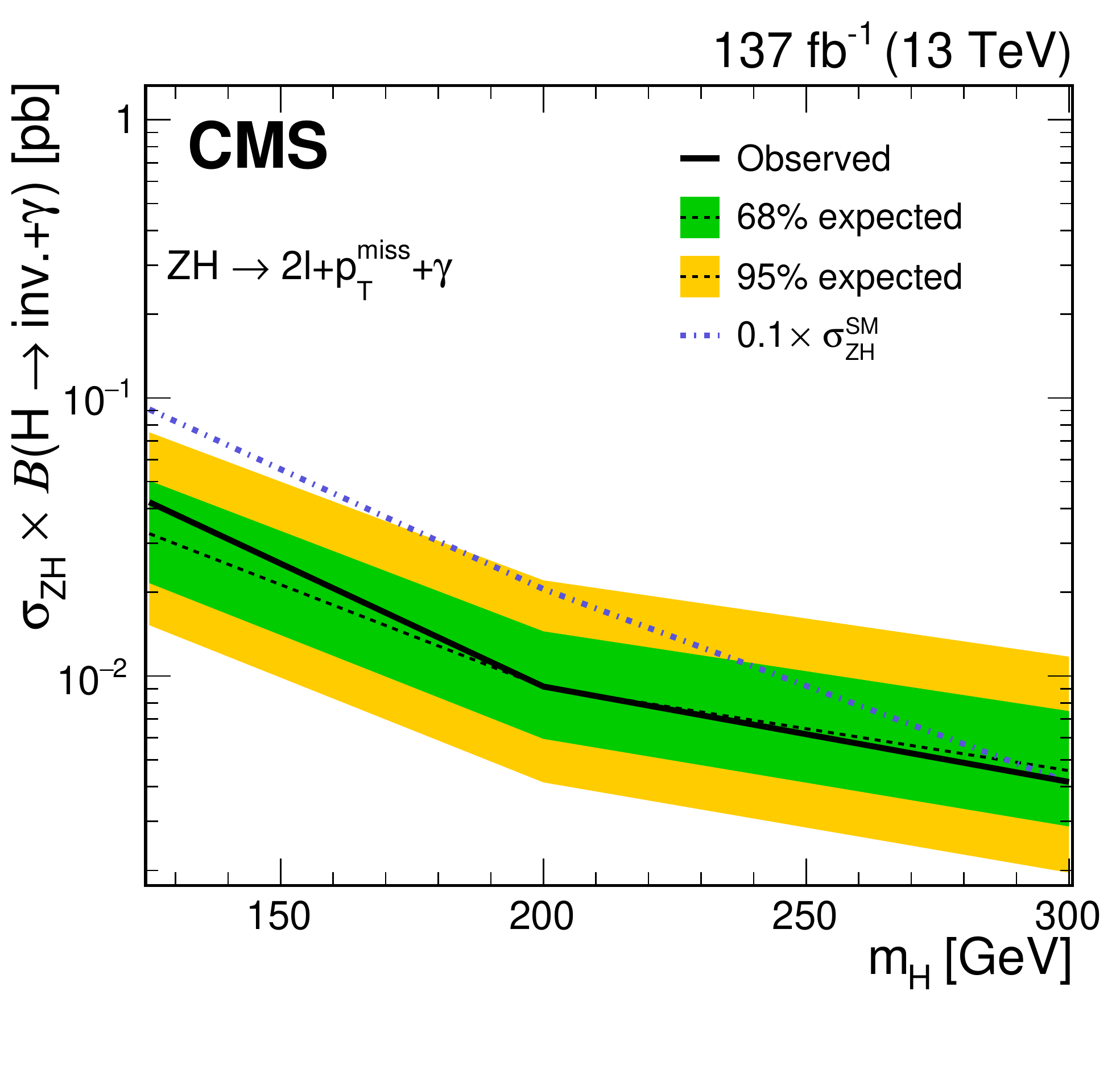}
 \caption{
   Expected and observed upper limits at 95\% $\CL$ on the product of 
   $\sigma_{\cPZ\PH}$ and $\mathcal{B}(\PH \to \text{invisible}+\gamma)$ as a 
   function of $m_{\PH}$. 
   The dot-dashed line shows the predicted  signal corresponding to 
   $0.1 \sigma_{\cPZ\PH}$.\label{fig:limits}}
\end{figure}

By using the fit strategy described in Section~\ref{sec:fitting}, 
upper limits as a function of $m_{\PH}$ are derived for the product of $\sigma_{\cPZ\PH}$ 
and $\mathcal{B}(\PH \to \text{invisible}+\gamma)$. 
For $m_{\PH} = 125\GeV$, this result can be interpreted as an upper limit on $\mathcal{B}(\PH \to \text{invisible}+\gamma)$ 
assuming the production rate for an SM Higgs boson~\cite{deFlorian:2016spz}. 
The upper limits at 95\% $\CL$ are calculated using a 
modified frequentist approach  with the $\CLs$ criterion~\cite{Read1,Junk:1999kv} 
and asymptotic method for the test statistic~\cite{LHC-HCG-Report,Cowan:2010js}. 
The observed (expected) 95\% $\CL$ upper limit at $m_{\PH} = 125\GeV$ on 
$\mathcal{B}(\PH \to \text{invisible}+\gamma)$ is 4.6 ($3.6^{+2.0}_{-1.2}$)\%. 
The expected and observed cross section upper limits at 95\% $\CL$ 
on the product of $\sigma_{\cPZ\PH}$ and $\mathcal{B}(\PH \to \text{invisible}+\gamma)$ as 
a function of $m_{\PH}$ are shown in Fig.~\ref{fig:limits}. 
Exclusion limits at 95\% $\CL$ on the product of  
$\sigma_{\cPZ\PH}$ and $\mathcal{B}(\PH \to \text{invisible}+\gamma)$ range from 
$\sim$40 to $\sim$4\unit{fb} as $m_{\PH}$ increases from 125 to 300\GeV. 
These limits also apply to other models where a scalar particle decays to a photon and light invisible particles.

\section{Summary}
\label{sec:summary}

A search is presented for a Higgs boson produced in association with a 
$\cPZ$ boson and decaying to an undetected particle together 
with an isolated photon. 
The analysis is based on a data set recorded by the CMS experiment in 2016--18 at a center-of-mass 
energy of $13\TeV$, corresponding to an integrated luminosity of $137\fbinv$. 
No significant excess of events above the expectation from standard model backgrounds is found. 
The results are used to place limits on the product of the cross section for 
associated $\cPZ\PH$ production and the branching fraction for such 
decays of the Higgs boson, in the context of a theoretical model where the undetected particle is a massless dark photon.
The observed and expected upper limits at 95\% confidence level at $m_{\PH} = 125\GeV$ on 
$\mathcal{B}(\PH \to \text{invisible}+\gamma)$, assuming standard model $\cPZ\PH$ associated 
production, are 4.6 and 3.6\%, respectively. 
Allowing for deviations from standard model $\cPZ\PH$ production, 
the product of $\sigma_{\cPZ\PH}$ and $\mathcal{B}(\PH \to \text{invisible}+\gamma)$ 
is excluded above $\sim$40 to $\sim$4\unit{fb}, 
for $m_{\PH}$ ranging from 125 to 300\GeV. 
These are the first limits on Higgs boson decays to final states that include an undetected massless dark photon.

\begin{acknowledgments}
We congratulate our colleagues in the CERN accelerator departments for the excellent performance of the LHC and thank the technical and administrative staffs at CERN and at other CMS institutes for their contributions to the success of the CMS effort. In addition, we gratefully acknowledge the computing centers and personnel of the Worldwide LHC Computing Grid for delivering so effectively the computing infrastructure essential to our analyses. Finally, we acknowledge the enduring support for the construction and operation of the LHC and the CMS detector provided by the following funding agencies: BMBWF and FWF (Austria); FNRS and FWO (Belgium); CNPq, CAPES, FAPERJ, FAPERGS, and FAPESP (Brazil); MES (Bulgaria); CERN; CAS, MoST, and NSFC (China); COLCIENCIAS (Colombia); MSES and CSF (Croatia); RPF (Cyprus); SENESCYT (Ecuador); MoER, ERC IUT, PUT and ERDF (Estonia); Academy of Finland, MEC, and HIP (Finland); CEA and CNRS/IN2P3 (France); BMBF, DFG, and HGF (Germany); GSRT (Greece); NKFIA (Hungary); DAE and DST (India); IPM (Iran); SFI (Ireland); INFN (Italy); MSIP and NRF (Republic of Korea); MES (Latvia); LAS (Lithuania); MOE and UM (Malaysia); BUAP, CINVESTAV, CONACYT, LNS, SEP, and UASLP-FAI (Mexico); MOS (Montenegro); MBIE (New Zealand); PAEC (Pakistan); MSHE and NSC (Poland); FCT (Portugal); JINR (Dubna); MON, RosAtom, RAS, RFBR, and NRC KI (Russia); MESTD (Serbia); SEIDI, CPAN, PCTI, and FEDER (Spain); MOSTR (Sri Lanka); Swiss Funding Agencies (Switzerland); MST (Taipei); ThEPCenter, IPST, STAR, and NSTDA (Thailand); TUBITAK and TAEK (Turkey); NASU and SFFR (Ukraine); STFC (United Kingdom); DOE and NSF (USA).

\hyphenation{Rachada-pisek} Individuals have received support from the Marie-Curie program and the European Research Council and Horizon 2020 Grant, contract Nos.\ 675440, 752730, and 765710 (European Union); the Leventis Foundation; the A.P.\ Sloan Foundation; the Alexander von Humboldt Foundation; the Belgian Federal Science Policy Office; the Fonds pour la Formation \`a la Recherche dans l'Industrie et dans l'Agriculture (FRIA-Belgium); the Agentschap voor Innovatie door Wetenschap en Technologie (IWT-Belgium); the F.R.S.-FNRS and FWO (Belgium) under the ``Excellence of Science -- EOS" -- be.h project n.\ 30820817; the Beijing Municipal Science \& Technology Commission, No. Z181100004218003; the Ministry of Education, Youth and Sports (MEYS) of the Czech Republic; the Lend\"ulet (``Momentum") Program and the J\'anos Bolyai Research Scholarship of the Hungarian Academy of Sciences, the New National Excellence Program \'UNKP, the NKFIA research grants 123842, 123959, 124845, 124850, 125105, 128713, 128786, and 129058 (Hungary); the Council of Science and Industrial Research, India; the HOMING PLUS program of the Foundation for Polish Science, cofinanced from European Union, Regional Development Fund, the Mobility Plus program of the Ministry of Science and Higher Education, the National Science Center (Poland), contracts Harmonia 2014/14/M/ST2/00428, Opus 2014/13/B/ST2/02543, 2014/15/B/ST2/03998, and 2015/19/B/ST2/02861, Sonata-bis 2012/07/E/ST2/01406; the National Priorities Research Program by Qatar National Research Fund; the Ministry of Science and Education, grant no. 3.2989.2017 (Russia); the Programa Estatal de Fomento de la Investigaci{\'o}n Cient{\'i}fica y T{\'e}cnica de Excelencia Mar\'{\i}a de Maeztu, grant MDM-2015-0509 and the Programa Severo Ochoa del Principado de Asturias; the Thalis and Aristeia programs cofinanced by EU-ESF and the Greek NSRF; the Rachadapisek Sompot Fund for Postdoctoral Fellowship, Chulalongkorn University and the Chulalongkorn Academic into Its 2nd Century Project Advancement Project (Thailand); the Welch Foundation, contract C-1845; and the Weston Havens Foundation (USA).
\end{acknowledgments}

\bibliography{auto_generated}
\cleardoublepage \appendix\section{The CMS Collaboration \label{app:collab}}\begin{sloppypar}\hyphenpenalty=5000\widowpenalty=500\clubpenalty=5000\input{EXO-19-007-authorlist.tex}\end{sloppypar}
\end{document}

%% file: EXO-19-007-authorlist.tex
\vskip\cmsinstskip
\textbf{Yerevan Physics Institute, Yerevan, Armenia}\\*[0pt]
A.M.~Sirunyan$^{\textrm{\dag}}$, A.~Tumasyan
\vskip\cmsinstskip
\textbf{Institut f\"{u}r Hochenergiephysik, Wien, Austria}\\*[0pt]
W.~Adam, F.~Ambrogi, T.~Bergauer, J.~Brandstetter, M.~Dragicevic, J.~Er\"{o}, A.~Escalante~Del~Valle, M.~Flechl, R.~Fr\"{u}hwirth\cmsAuthorMark{1}, M.~Jeitler\cmsAuthorMark{1}, N.~Krammer, I.~Kr\"{a}tschmer, D.~Liko, T.~Madlener, I.~Mikulec, N.~Rad, J.~Schieck\cmsAuthorMark{1}, R.~Sch\"{o}fbeck, M.~Spanring, D.~Spitzbart, W.~Waltenberger, C.-E.~Wulz\cmsAuthorMark{1}, M.~Zarucki
% ...
\vskip\cmsinstskip
\dag: Deceased\\
1:  Also at Vienna University of Technology, Vienna, Austria\\
% ...
